# Ultra-flexible silicon foils with seamless detachability: the effect of porous multilayered structures prepared through modulated electrolyte composition.

Authors:

C. Sanchez-Perez[1,2] – email: clara.sanchez.perez@upm.es, clara.sanchez@urjc.es
P. Rivas-Lazaro[1] – paula.rivas@alumnos.upm.es
E. García-Tabarés[3] – egarciat@fis.uc3m.es
I. García[1]* (corresponding author) – ivan.garciav@upm.es

1. Instituto de Energía Solar, Universidad Politécnica de Madrid (IES-UPM), Av. Complutense s/n, 28040 Madrid, Spain
2. Present address: Chemical and Environmental Engineering Group, ESCET, Universidad Rey Juan Carlos, C/Tulipán s/n, Móstoles, Madrid 28933, Spain.
3. Physics Department, Universidad Carlos III de Madrid (UC3M), Av. Universidad 40, 28911 Leganés, Spain.

## Abstract

A comprehensive evaluation of the effect and limitations of variable current density and electrolyte composition on layer porosity and microstructure changes of porous silicon (pSi) multilayer stacks is reported. Following these results, the development and optimization of a four-layer stack architecture is reported through addition of super-low porosity layers (SLPL) on a low/high porosity layer (LPL/HLP) stack. Thermal treatment of these structures achieved excellent top and bottom surface reconstruction to form sintered porous silicon (SPS) detachable foils, enabling direct foil separation using cello tape from a smooth and specular parent substrate without the need to cut or damage it.

## 1. Introduction

Silicon (Si) is the most popular material used in electronics and optoelectronics due to its semiconducting properties, its natural abundance, and its competitive price. The development of thin and ultra-thin Si foils has marked the future of flexible optoelectronics, particularly in the field of photovoltaics.[1,2] In addition, ultra-thin Si foil fabrication enabled the possibility of substrate reuse after foil detachment, which results into a further advantage as 30% of the overall expenses for a silicon-based solar cell are attributed to the silicon material.[3,4]

There is a spectrum of standardized methods to prepare thin Si foils,[5] which are flexible but also brittle. Instead, porous Si (pSi) membranes have high flexibility and elasticity but are prone to oxidation. Alternatively, sintered porous silicon (SPS) foils can be synthesised with oxide-free and highly flexible structures from double-layer stacks, featuring an inner porous core capped with a 10 nm-thick void free surface layer.[6,7] These substrates have enormous potential in the field of microelectronics, and have structural, thermal and optical properties very attractive for photovoltaic applications,[8] *i.e.* their potential as virtual substrate for III-V/SPS integration due to its tuneable lattice constant.[9–11]

Free-standing ultra-thin Si foils are typically prepared *via* electrochemical etching of tubular macropores patterned through photolithography [12,13] followed by sintering at temperatures usually above 1000 ˚C.[14] However, they involve expensive and non-scalable steps, limit foil thickness to 1-2 μm and exhibit issues regarding foil detachability. Electrochemical etching of meso-pSi using a HF aqueous-based electrolyte has been researched as an alternative to prepare a wider thickness range, offering a more affordable approach with a wider synthesis window in terms of pore size/type and porosity tunability. Variations on the applied current density (*j*) or the electrolyte concentration offer easy control over pore size and type during a unique electrochemical process.[15] With this method, the production of bilayers of low porosity/high porosity (LPL/HPL) sintered at high temperature in a reducing atmosphere has yielded easily detachable Si foils,[16–18] and subsequent Si epitaxy has been used to prepare Si ultra-thin foils integrated in photovoltaic applications.[3] To this date, standalone SPS layers still do not hold potential for optoelectronic integration due to the formation of surface defects such as open voids,[6] as well as the formation of relatively thick bonding pillars between the foil and the parent substrate.[13] The annealing temperature to produce SPS foils with a porous core is also limited, as very high sintering temperatures may produce complete reorganization of the porous core to form a solid layer, decreasing its flexibility. In addition, it has been reported that generally high temperature increases surface roughness, worsening foil quality as epitaxy seed.[19] Long sintering times produce smoother top reconstruction, but a trade-off between these parameters must be reached to obtain epi-like foil surfaces in a reasonable timeframe.

Recent research carried out by the authors of this paper highlighted the benefits of LPL/HPL stack formation using low current density and variable electrolyte composition to prepare SPS foils with very low surface roughness, as well as easy foil detachment through the development of the HPL into a pSi "mesh" rather than an arrangement of pillars or a simple void.[9] Although the proposed methodology enables parent substrate reuse, considerable reconditioning is still necessary for it. Another way to increase surface reconstruction with low roughness has been proposed adding very thin PSi layers, either to enhance top quality[6,16] or as an intermediate layer between the LPL and the HPL to enhance foil separation.[20] Following both approaches, in this contribution we propose the design of a multilayer pSi stack to create a SPS foil with fully sealed and smooth top and bottom surfaces through "seamless" foil detachment, leaving a parent substrate with very low RMS facilitating its reconditioning for further reuses. This is achieved through the introduction of "super low" porosity layers (SLPL) in the structure, which have been reported to provide very smooth reconstructed surfaces when prepared using low current density and high electrolyte concentration.[21] This design comprises a four-layer structure with a SLPL top layer, followed by a layer with slightly higher porosity, *i. e.* LPL, another SLPL and a bottom HPL.

## 2. Experimental

### 2.1. Synthesis and characterisation

Porous silicon layers (1.8 cm$^2$ and 19.6 cm$^2$) were etched on diced single-side polished boron doped Si(100) oriented wafers with a 6˚ miscut towards the (111) plane and a resistivity of 0.01 Ωcm. Wafer cleaning and electrochemical experiments were carried out as previously described elsewhere.[9] Electrolyte [HF(50%):EtOH(99%)] solutions were prepared with ratios ranging between 1:0 - 1:4 (25-10 w%). Etching was performed at a constant *j* in the range of 5 - 90 mA/cm$^2$, using a Keithley 2602 precision current source. PSi samples were sintered in reducing conditions at 950 ˚C for 1 h, as described elsewhere.[9] High resolution X-ray diffraction (HRXRD) Panalytical X'PERT diffractometer with monochromated Cu Kα1 radiation (1.54184 Å, 45 kV, 30 mA) equipped with an Eulerian cradle suitable for both double axis (rocking curve) and triple axis (coupled scan) configurations, with which diffraction intensity vs. rocking angle patterns were collected. The crystal quality of samples was evaluated on the (004) silicon diffraction peak. Micrographs were collected with a JEOL JSM 6301F Field Emission SEM (3 keV, SEM mode, WD = 8 mm). Surface roughness was characterized using an AFM Park XE-10 operated in tapping mode, averaging 3-9 measurements per sample. Layer thicknesses and pore sizes were calculated using ImageJ software averaging data from at least 3 points for thicknesses and 20 features per micrograph. Sample porosity was evaluated through a pixel-by-pixel analysis of SEM micrographs using an ad-hoc made Python-based program where pixels were categorized by colour as either a wall/interpore (bright pixels) or a pore (dark pixels) following a presser brightness threshold. This threshold was based on a comprehensive analysis of SEM micrographs of a selection of monolayers reproduced from literature, with which porosity values were adjusted. The calculation of sample porosity was carried out with Eq. 1, after which the program generates an image in which pores are highlighted in blue as a visual aid to ensure a good performance of the algorithm (Fig. S1).:

$$P(\%) = \frac{N_{pore\ pixels}}{N_{pore\ pixels} + N_{wall\ pixels}} \qquad \text{Eq. 1}$$

## 2.2. Experimental design

Etching parameters (current density $j$, electrolyte concentration HF(%), time) affect the porous microstructure of an etched (E) bi/multilayer. Sample porosity P(%) is commonly reported in SPS literature, as indication of its tendency to increase/decrease porosity during sintering.[18] However, this value holds very limited information since samples with similar porosity values P(%) can have strikingly different pore types and pore/interpore aspect ratios.[9] Thermally activated pore reorganization will be affected by initial porous structure, so the microstructure of sintered (S) bi/multilayers can be drastically different for two samples with similar porosity but different microstructure. Therefore, the fabrication of SPS detachable foils requires a deep insight on the effect of each synthetic parameter involved to optimize both etching and sintering steps to the point of allowing the development of reproducible synthetic recipes.

Pore size enlargement is directly related to increases in $j$ at a given HF(%) (from here on referred to as method A) or a decrease in HF(%) at a given $j$ (from here on referred to as method B), both leading to the formation of PSi layer with larger pores.[22] The dissolution reaction for each method is triggered by different reaction mechanisms, so each resulting porous surface can have different exposed facets, which would translate into energetically different starting points for thermally-driven reorganization.[23]

Si atomic diffusion is thermally driven to lower the surface free energy for temperatures higher than 700 ˚C,[7] so pore reorganization can lead to either pore enlargement or collapsing depending on the pore immediate surroundings. Porous monolayers with P(%) < 30% close up the porous structure, while in layers with P(%) > 70 % pore size is driven to increase through pore coalescence.[24] In multilayer systems, adjacent layers with different P(%) affect each other´s reorganization, so in LPL/HPL bilayers the pores of the LPL tend to collapse whereas pores in the HPL tend to open.[7] Therefore, the design of a LPL/HPL interface will depend on both layer porosity and the difference in porosities of each layer. For instance, sintering of a LPL/HPL bilayer with similar P(%) values will lead to the formation of a LPL with lower P(%) and a HPL with higher P(%) than in the etched sample, having an interface region in which porosity varies gradually. On the other hand, sintering of LPL/HPL stacks in which layers have very different P(%) values will promote enhanced pore collapse in the LPL and pore coalescence in the HPL, which can form a "void" if the difference of porosity between layers is extreme.

The LPL/HPL stack approach has long been employed to fabricate detachable μm-thick silicon foils but have proven difficult to ensure satisfactory top and bottom surface reconstruction to form a featureless "void" for detachment. Ideally, the top surface quality should be in the range of epi-ready wafers (RMS ~ 0.2 nm) to be suitable for opto-electronic applications, and the foil should be easily detachable without damage, which depends on the structural characteristics of that "void". For instance, if there are no connecting pillars between the foil and the parent substrate the foil can collapse and break,[14] but if the pillars are wide, they do not break easily and then foils do not detach.[16,25] Therefore, sintering of the LPL/HPL interface needs to be carefully tuned to avoid both situations. For that we propose a novel design that involves the addition of a "super low porosity layer" (SLPL) in each interface of the LPL.

Firstly, the addition of a SLPL over the LPL acts as a LPL/HLP stack with low P(%) that should enhance pore closure to force a fully reconstructed surface while retaining a porous core, and a gradual porosity between layers. The closed surface protects the porous core from oxidation, forming a thin barrier layer of nm-sized oxide in contact with air that is removed during sintering, which has been proven to lead to very low surface RMS (Fig. 1a-b).[9] Then, the addition of a second SLPL over the HPL acts as LPL/HPL with extreme P(%) difference, enhancing reconstruction of the bottom surface of the foil to form very loosely attached pillars on the "void" (Fig. 1c-d). Tuning the structural parameters of the porous layers involved in this interface is expected to lead to "seamless" detachment. Hence, the proposed novel multilayered architecture based on varied LPL/LPL interfaces is expected to simultaneously improve top surface quality for further epitaxy and foil detachability.

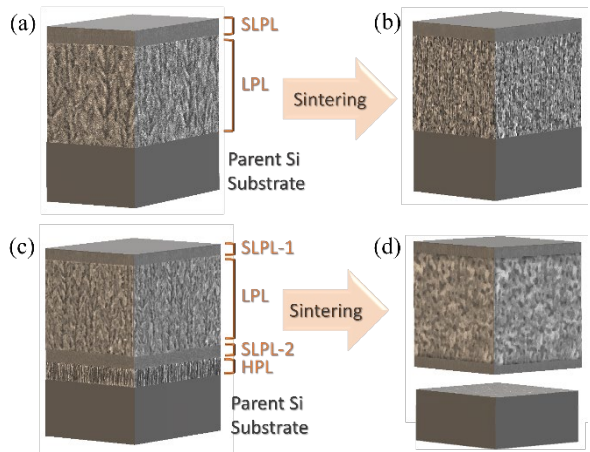

**Figure 1.** 3D sketches of porous stacks etched over a (001) Si substrate, namely (a) as-etched and (b) sintered SLPL/LPL bilayers, and (c) as-etched and (d) sintered SLPL/LPL/SLPL/HLP multilayer stacks.

HF(50%) aqueous electrolytes are widely used to produce porous Si due to the very high dissolution rates achieved. However, such electrolyte is unsuitable to fabricate multilayers with thick LPLs. This is because pore formation is accompanied by $H_2$ evolution, which cannot diffuse through the SLPL and remains trapped inside the branched structure building up pressure and ultimately breaking the porous structure (Fig. S2). Therefore, only ethanol-based HF electrolytes were employed to obtain reproducible results.

## 3. Results and discussion

### 3.1 PSi bilayers

The microstructure, P(%) and thickness of each porous layer sets the initial conditions to optimize sintering protocols. Therefore, the use of different synthesis methods to etch bilayers may lead to different structures. The first section of this work seeks to evaluate the effect of a top SLPL on the surface reconstruction of a pSi layer, for which two bilayer (BL) samples of 5 $\mu m$-thick SLPL/LPL stacks were etched on Si wafers and sintered in identical conditions. The SLPLs were prepared identically for both samples ($j$ = 5 mA/cm$^2$; HF(%) = 50%, thickness ~300 nm) and then LPLs were etched either using method A (sample BL-AE) or B (sample BL-BE) (Table 1).

HRXRD analysis of the bilayer samples (Fig. 2a and 2b) reveals fundamentally different LPL microstructures. It can be observed that in both samples pSi diffraction peaks appear shifted 0.045 degrees from the Si peak, indicating similar out-of-plane residual strain in both porous structures. Reflections were collected using double axis (DA) and triple axis (TA) configurations. A similar envelope is observed for TA measurements of both samples, but a much wider envelope is observed for sample BL-AE when the DA configuration is employed that indicates the existence of tilted crystalline planes in the microstructure of the porous layer. This result is consistent with the SEM micrographs that show a branched microstructure with high mosaicity in sample BL-AE (Fig. 2c) and a heavily columnar orientation of pores with low mosaicity in sample BL-BE (Fig. 2d). Hence, pore morphology and pore interpore/wall sizes (Table 1) are in good agreement with the virtually identical residual strain generated by the LPL for etched samples.

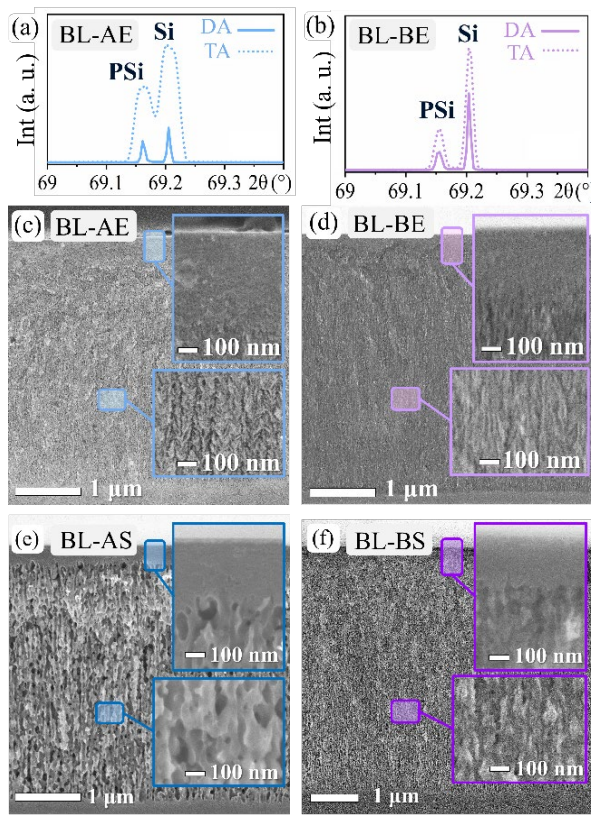

**Figure 2.** HRXRD patterns collected in double axis (DA) and triple axis (TA) configurations of 5 μm thick bilayers etched with (a) method A and (b) method B, showing the diffraction peak from Si wafer (Si) and the diffraction peak that appears from the out-of-plane tensile stress of the porous layer (pSi). Cross-sectional SEM micrographs of the etched samples (c) BL-AE and (d) BL-BE. Cross-sectional SEM micrographs of 5 μm etched and sintered bilayers (e) BL-AS and (f) BL-BS.

**Table 1.** Summary of thickness, current density ($j(mA/cm^2)$), electrolyte concentration ($[HF](\%)$) sintering conditions and foil detachability, porosity P(%), as well as pore (P) diameter and wall or interpore (IP) diameter of the porous embedded foil core (LPL) of studied samples. Standard deviation of diameter measurements (s.d.) in indicated in brackets.

| Sample ID | Thick (μm) | Synthesis parameters ||||  Sinter/Detach 950 °C 1 h | LPL porosity P(%) (%) | LPL pore diameter (s.d.) (nm) | LPL interpore diameter (s.d.) (nm) |
|---|---|---|---|---|---|---|---|---|---|
| | | LPL || HPL || | | | |
| | | $j$ | $[HF]$ | $j$ | $[HF]$ | | | | |
| **BL-AE** | 5 | 50 | 25 | - | - | No/No | 22 | 13 (2) | 20 (4) |
| **BL-BE** | 5 | 10 | 17 | - | - | No/No | 14 | 15 (3) | 15 (3) |
| **BL-AS** | 5 | 50 | 25 | - | - | Yes/No | 28 | 80 (20) | 80 (20) |
| **BL-BS** | 5 | 10 | 17 | - | - | Yes/No | 24 | 21 (5) | 16 (3) |
| **ML2-A** | 2 | 50 | 25 | 70 | 25 | Yes/Yes | 30 | 54 (12) | 45 (9) |
| **ML2-B** | 2 | 10 | 17 | 10 | 12 | Yes/Yes | 50 | 14 (55) | 250 (100) |
| **ML10-A** | 10 | 50 | 25 | 90 | 25 | Yes/Yes | 19 – 35 | 62(16) –249(50) | 83(19) – 109(37) |
| **ML10-B** | 10 | 10 | 17 | 20 | 12 | Yes/Yes | 35 – 40 | 41(12) – 69(27) | 52(12) – 78(13) |

However, a higher porosity was obtained for sample BL-AE (22.4%) than for sample BL-BE (13.7%). These differences in P(%) and microstructure between etched samples show how the selected etching method will have a great effect over thermal reorganization. To evaluate this effect quantitatively, another two bilayers were etched identically and sintered at 950 °C in $H_2$ for 1 hour (samples BL-AS and BL-BS).

The SLPL/LPL interface of both samples behaved as expected for a LPL/HPL stack with low P(%) difference between layers. SEM micrographs of sintered bilayers (Fig. 2e-f) confirmed that sintering prompted pore collapsing on the SLPL and pore merging and reconfiguration on the LPL.

Although they both exhibit qualitatively similar structures based on elongated hourglass-shaped pores, they have very different pore and pore wall sizes. Sample BL-AS featured a small porosity enhancement alongside a large increase of pore size and the disappearance of the branched structure while sample BL-BS retained pore and wall sizes undergoing a larger porosity enhancement (Table 1). Overall, an enhanced pore reorganization was observed in sample BL-AS, a result consistent with the larger P(%) difference between SLPL and LPL in the etched microstructure.

Regarding top surface quality, pore formation on a wafer surface increases surface roughness, and then sintering is expected to "close" the surface, reducing its roughness. AFM analysis of the top surface of the bilayer samples (Fig. 3) showed average RMS values for etched samples comparable to those of epi-ready Si (001) wafers with 6º miscut (Fig. 3a, 3c). Such low RMS value for electrochemically etched samples is not unexpected considering that the pores of a SLPL ought to be below 5 nm. Another benefit of having surface pores within this range is that they favour the formation of a thin top native oxide "crust" that protects the porous structure from oxidation.[9] Surface roughness of sintered samples increased for the sample etched with high current density (Fig. 3b) whereas it remained low for the one etched with high dilution (Fig. 3d). This difference could be related to the relative silicon mobility in exposed facets during diffusion.[23] It is worth noting that despite the relatively low RMS values of all four samples, worse surface quality due to appearance of needle-like extrusions are observed in 3D maps (inset images) of samples prepared *via* method A (Fig. 3a, 3b). The reported RMS values for sintered samples are significantly lower than those found in the literature, [6,16–18] which provides evidence that adding a top SLPL increases surface reconstruction and minimizes top surface roughness. What is more, despite the increase in RMS resulting from the electrochemical and thermal treatment, surface quality lies above that reported adequate for subsequent epitaxial growth for sample BL-BS.[26]

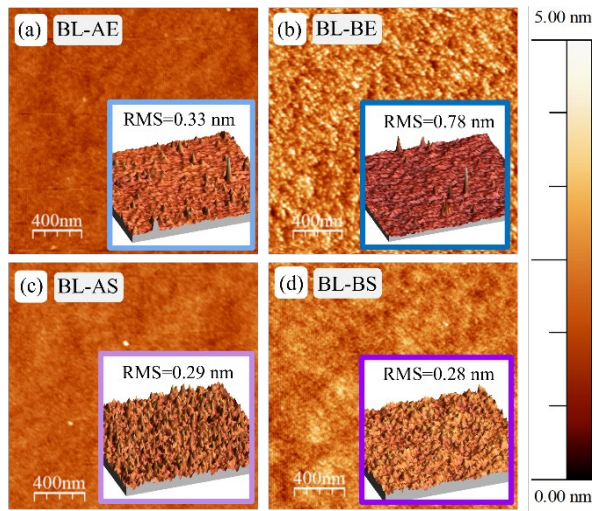

**Figure 3.** AFM measurements of bilayer samples prepared though method A namely (a) etched and (b) sintered, and samples prepared though method B namely (c) etched and (d) sintered.

3.2 PSi multilayers

The multilayer stack design consisted of a SLPL/LPL/SLPL/HPL architecture (Fig. 1 c). To study the effect of varying synthetic method and film thickness on the sintered microstructure, four different multilayer (ML) samples were etched and sintered. Their thicknesses were selected as representative minimal thicknesses for thin and ultra-thin foils, considering multilayer detachable stacks are made of 3 layers, as well as acceptable time constrictions (below 1 hour).

SLPLs were prepared as previously described in all cases. ML Samples were labelled indicating their final thickness (2 or 10 μm) and the synthesis method used for the LPL and HPL layers (A or B), namely ML2-A, ML10-A, ML2-B and ML10-B (Table 1). Pore development of the different layers of these samples during sintering proceeded as expected: the low porosity difference between

SLPL and LPL (foil top) induced a smooth porosity gradient upon sintering of the SLPL/LPL interface, favouring closing-up of the top surface of the SLSP over which a thin layer of native oxide (roughly 10-15 nm) is formed upon exposure to air (Fig. S3). In turn, the large difference in porosity between the SLPL and the HPL in the SLPL/HPL interface (foil bottom) led to the formation of a closed surface over a void in optimal conditions. This occurred when the thickness of the HPL was smaller than the size of sintered pores and made the PSP foil detachable.

Cross-sectional SEM images of ML samples are shown in Figure 4. It was found that (a) SLPLs over 500 nm thick at the SLPL/HPL interface induced optimal reorganization to maximize porosity gap between the SLPL and the LPL and (b) optimal HPL thickness was found at approximately 50 nm, producing connecting pillars near 50 nm thick in all cases. Following this result, multilayered detachable foils were successfully prepared for 2 $\mu m$ (Fig. 4a) and 10 $\mu m$ (Fig. 4b) total thickness.

A notable effect of method A is that the sintered HPL forms an array of very thin and fragile pillars that connect the foil and the substrate, which left a "hairy" surface after detachment (Fig. 4 insets). Additionally, a widely reported factor on pSi formation is that pore size typically increases with layer width due to limited diffusion of HF through the pores at high etching rates (see table 1).[22] This effect is particularly relevant for thick MLs etched using method A, and it can be observed in the acute porosity gradient (16% variation from top to bottom) of the sintered LPL in Fig. 6d. Consequently, the larger P(%) difference of the LPL and SLPL in the LPL/SLPL/HPL region of sample ML10-A leads to a weakly bonded SLPL (Fig. 4b).

In contrast, the etching process using method B has a slower dissolution rate, which allows continuous HF diffusion through the pores. Hence, it produced a homogeneous porosity profile (4.8% variation from top to bottom) even for thick foils (Fig. 4d), making both SLPL/LPL top and bottom interfaces structurally strong. Regarding HPL design, sintering of ML samples gave optimized oval-shaped pores of up to 400 nm average diameter with thin connecting pillars for HF (10%) electrolyte concentration (Fig. S4). Even though the size of connecting pillars is virtually equal in all four ML samples, the high aspect ratio between pore and pillar sizes in the LPL forms a smoother back surface for thin foils (Fig. 4c) and almost seamless detachment of thick foils (Fig 4d).

Form the observed etched and sintered structures it can be concluded that the use of high current densities (method A) increases layer porosity due to the formation of ramified pores, while doing it *via* electrolyte dilution (method B) occurs by forming wider pores. Then, it is reasonable that sintering induces the formation of a dense array of pillars in the first case and disperse assortment of pillars in the second one.

Regarding the quality of the top foil surface AFM analysis of the top foil surface and the parent substrate (after detachment of the Si foil) of ML samples is shown in Figure 5. It can be observed that RMS values for top foil surfaces prepared *via* method A exhibit a smoother RMS baseline but with a significant number of "events" (Figs. 5a, 5c) that increase surface roughness. Instead, method B leads to a lower degree of surface reconstruction and a slightly higher general roughness but with limited "events" (Fig. 5b, 5d). Therefore, the overall average RMS of ML samples etched with method A is roughly 0.3 nm higher than in case of method B for both thicknesses (Table S1).

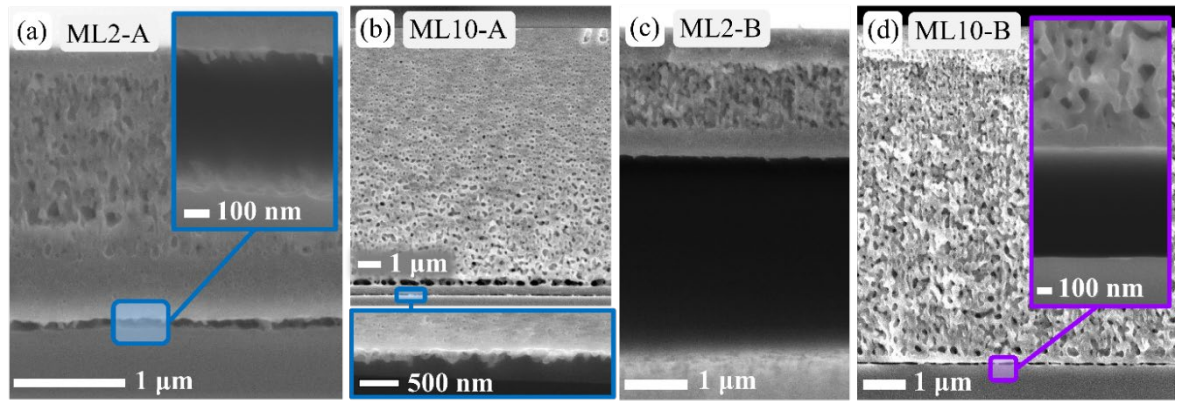

**Figure 4.** Cross-sectional SEM micrographs of etched and sintered detachable multilayers for thickness (a) 2 *μ*m and (d) 10 *μ*m using method A, and (c) 2 *μ*m and (d) 10 *μ*m using method B. SEM insets in (a) and (d) are tilted 5 º.

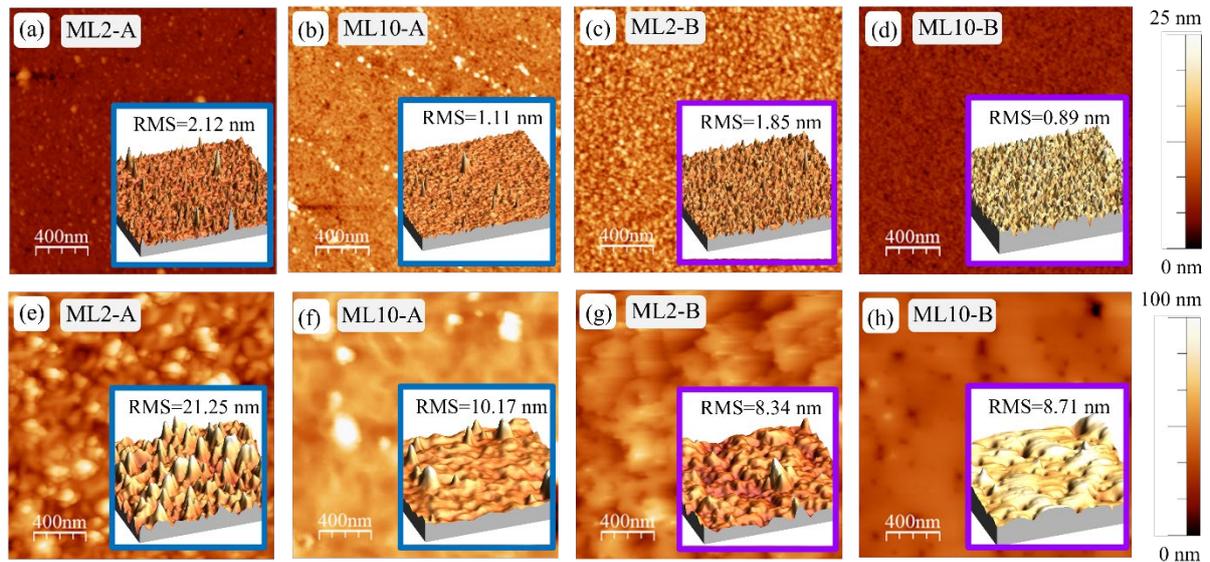

**Figure 5.** AFM of foil top surface (a-d, top row) and parent substrate after detachment (e-h, bottom row) of free-standing sintered multilayers prepared via method A or B with thicknesses 2 and 10 *μ*m.

It is notable that RMS values for 2 μm thick MLs remain at ~2 nm whereas those for 10 μm are close to 1 nm. No direct comparison of the RMS values of the reconstructed surfaces of bilayers (Fig. 3b, 3d) and of the top surface of free-standing foils (Fig. 5a-d) would be scientifically relevant: bilayers are not detachable so higher silicon diffusion in the LPL occurs towards the surface, while silicon diffusion in the LPL of multilayers occurs both towards the top surface and the bottom surface. Hence, a lower degree of top reconstruction is expected in multilayer structures. Still, the use of a top SLPL provides a smooth reconstructed surface with RMS < 1nm, which hold potential for optimization to achieve surface quality for subsequent epitaxy upon optimization. It also promotes enhanced reorganization of the porous core in thicker foils, likely due to a larger mobility path for Si from the LPL to the top SLPL.

Surface analysis of the parent substrate after foil detachment shows again higher inhomogeneity in samples etched with method A, characterized by a greater number of "events" than in comparable samples etched with method B (Fig. 5e-h). Again, the number of "events" decreases with foil thickness, and it is interesting that those "events" found on the parent substrate after detachment of sample ML10-B are indentations rather that protrusions.

Optimization of synthesis parameters indicated that to obtain readily detachable foils without foil damage, both the thickness of the pillars attaching said foil to the parent substrate and the void width must be below 100 nm. That being said, optimal synthesis conditions using method A produce very rough surfaces after detachment (Fig.6a, b), which are not convenient for further contact deposition (bottom of foil) or require significant reconditioning for subsequent ML fabrication (top of parent surface after detachment). In turn, optimal conditions for method B produce significantly smoother surfaces on the parent substrate after detachment (fig. 6d), which are close to optimal to avoid substrate reconditioning altogether.

An example of our advanced ML method to produce freestanding oxide-free foils can be observed in Fig. 6, which were readily detached using Kapton tape. Synthesis conditions for sample ML10-A resulted on a surface of the parent substrate after layer detachment with a very rough surface (Fig. 6e), while a significantly smoother and mirror-like parent substrate surface was obtained for sample ML10-B (Fig. 6f). The porous LPL structure provided both foils with high flexibility (Fig. 6g-h), but the strikingly different reconstruction of the HPL lies on the fact that ¨large¨ pores grow wide (method B) rather than long (method A). To test this concept, a ML with the same structure than sample ML10-B was electrochemically etched and separated from the parent substrate without sintering, only using standard cello tape and without cutting/damaging the substrate. Free-standing etched ML foils of 1.8 cm$^2$ (Fig. 6i) and 19.6 cm$^2$ (Fig. S5) were successfully separated from the parent substrate, and exhibited a dark red tone expected due to light quantum confinement in the porous LPL structure (>10 nm thick walls).

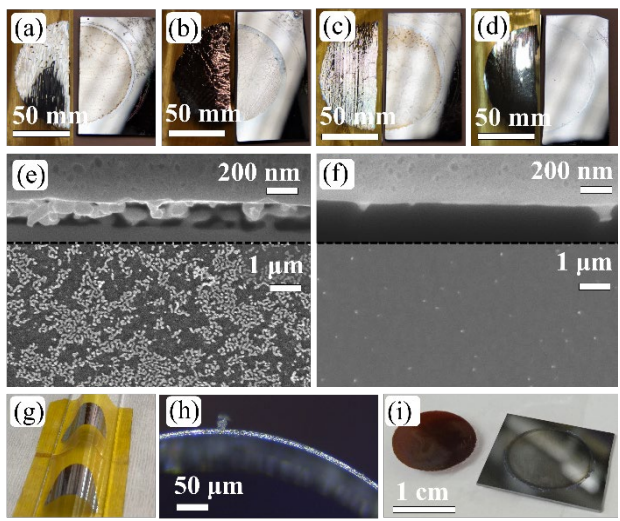

**Figure 6.** Photographs of foil back and parent substrate of ML sintered samples after detachment (a) ML2-A, (b) ML10-A, (c) ML2-B and (d) ML10-B. Cross-sectional SEM images of the HPL interface (top) and parent substrate surface (bottom) of samples (e) ML10-A and (f) ML10-B. Images showing high flexibility of ML detached using Kapton tape, both (g) macroscopic and (h) microscopic. (i) Picture of 10 μm-thick ML prepared using method B showing excellent detachment from parent substrate before sintering, so it can be readily detached using cello tape without cutting or damaging the parent substrate.

The parent substrate exhibited a featureless almost specular surface right after detachment. This confirmed that HPL formation through method B produces "seamless" foil detachment, indicating that standard cello tape can be used to detach sintered foils leaving a substrate surface in excellent conditions for minimal reconditioning for subsequent reuse (Fig. 7). Videos of the detachment process and flexibility tests are provided as supplementary material.

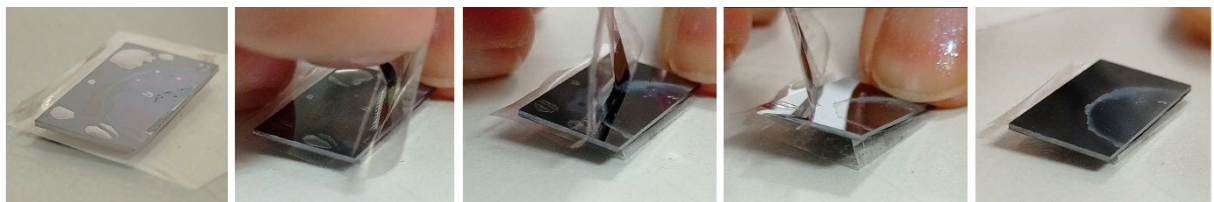

**Figure 7**. Photograph sequence of foil detachment from sample ML10-B using commercial cello tape, showing a flat and specular substrate surface.

## 4. Conclusions

The results presented herein provide valuable insight regarding the development of thin foil Si technology. The advanced multilayer porous stacks provide a deep insight of the interconnected effect of adjacent porous layers considering their different porosity and crystal configuration. Optimized fabrication of oxide-free Si foils is reported featuring (a) high flexibility due to their porous core, (b) excellent detachability leaving a smooth and specular parent substrate without cutting it and (c) a highly controllable and reproducible synthesis process suitable for easy scale-up. The main conclusions of this work can be listed as follows:

1. The interface between porous layers with different porosities produce a gradual P(%) gradient for $\Delta P(\%) < 40\%$ and a sharp P(%) change for $\Delta P(\%) > 40\%$. Voids for "seamless" foil detachment can be obtained for $\Delta P(\%) > 90\%$.

2. A "superlow" porosity layer (SLPL) adjacent to a LPL or HPL can be employed to protect the porous structure from oxidation upon exposure to air as well as to maximize surface reconstruction during sintering. Gradual and sharp P(%) gradients in SLPL/LPL and SLPL/HPL interfaces, respectively, ensure full top surface reconstruction and bottom void formation.

3. Regarding the fabrication of porous multilayers, it can be concluded that different microstructures are obtained upon choice of method, *i. e.* branched (method A) and mainly longitudinal (method B). Homogeneous P(%) depth dispersion in LPLs is only observed on samples prepared *via* method B, likely due to the better electrolyte diffusion given its lower etching rate.

4. Pore enlargement in the HPL depends on initial etched microstructure, affecting foil detachability and the quality of the detached surfaces. Longitudinally enlarged pores are formed using method A due to increased dissolution at the pore tip, which leaves "hairy" surfaces covered in small broken pillars upon detachment. Method B allows pores to grow also laterally, forming a void with lower density of pillars that enables "seamless" foil detachment. In all cases, optimal HPL thickness has been obtained for 100 nm thick layers for the selected parameters.

While the use of high current densities in method A (variable current density, high electrolyte concentration) can significantly speed up the etching process, it has proven incompatible with SLPL in foils thicker than 2 $\mu$m. They feature weak core porous structures due to large P(%) variations in the LPL, and they leave nanostructured "hairy" detached surfaces. In addition, it has important scalability drawbacks, since increasing the wafer's surface area requires the use of excessively high currents, which poses a challenge in attaining a perfectly uniform high current over the whole area. Moreover, the high currents handled can represent an operational risk and require costly instrumentation capable of providing the high currents with precision.

Method B (low current density, variable electrolyte concentration) is fully compatible with the presence of SLPLs, as it favours $H_2$ evolution through the porous structure. Its lower etching rate makes it more controllable and reproducible and optimal detachment has been achieved using this method, as both etched and sintered foils can be separated from the parent substrate using commercial cello tape. This is particularly relevant as there is no need for wafer cutting to obtain a free-standing foil and minimal substrate reconditioning is required, so subsequent processes could be performed to maximize wafer use. The novelty of this work also resides in the potential to design protocols including variable electrolyte concentration and 4-layer structures to improve the design of detachable foils of other related materials, *i. e.* Ge and GaAs.

*CRediT authorship contribution statement*

**Clara Sanchez-Perez**: Conceptualization, Methodology, Investigation, Validation, Formal analysis, Resources, Writing – original draft, Visualization, Project administration. **Paula Rivas-Lazaro**: Methodology, Validation, Investigation, Writing. **Elisa García-Tabarés**: Methodology, Formal analysis, Validation, Writing. **Ivan Garcia**:

Conceptualization, Resources, Investigation, Writing – review & editing, Supervision, Project administration, Funding acquisition.

**Declaration of Competing Interest**

The authors declare that they have no known competing financial interests or personal relationships that could have appeared to influence the work reported in this paper.

**Acknowledgements**

The authors are indebted to Luis Cifuentes and Jesús Bautista for their unvaluable technical assistance. This work has been supported by grant PID2021-123530OB-I00 funded by MICIU/AEI/ 10.13039/501100011033 and FEDER, UE.

**Appendix**

Electronic supporting information is available with a representation of the method employed to calculate layer porosities (Fig. S1), pictures and micrographs of samples obtained throughout the iterative optimization process that illustrate methodological limitations (Fig. S2). A table including average values and standard deviations of RMS measurements for foil surfaces and parent substrate surfaces of all samples (Table S1). Videos illustrating foil detachment, and the high flexibility of foils are also attached (Videos S1 and S2)

**Appendix A. Supplementary material**

Supplementary data to this article can be found online at xxx

# SUPPLEMENTAL INFORMATION

Ultra-flexible silicon foils with seamless detachability: the effect of porous multilayered structures prepared through modulated electrolyte composition.

Authors:

C. Sanchez-Perez[1,2] – email: clara.sanchez.perez@upm.es, clara.sanchez@urjc.es
P. Rivas-Lazaro[1] – paula.rivas@alumnos.upm.es
E. García-Tabarés[3] – egarciat@fis.uc3m.es
I. García[1]* (corresponding author) – ivan.garciav@upm.es

*1. Instituto de Energía Solar, Universidad Politécnica de Madrid (IES-UPM), Av. Complutense s/n, 28040 Madrid, Spain*
*2. Present address: Chemical and Environmental Engineering Group, ESCET, Universidad Rey Juan Carlos, C/Tulipán s/n, Móstoles, Madrid 28933, Spain.*
*3. Physics Department, Universidad Carlos III de Madrid (UC3M), Av. Universidad 40, 28911 Leganés, Spain.*

## ELECTRONIC SUPPORTING INFORMATION

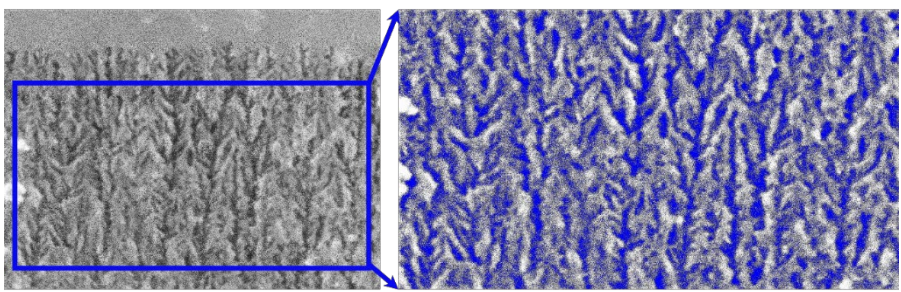

**Figure S1.** Representative image of pixel-by-pixel analysis to calculate layer porosity using the developed python-based program.

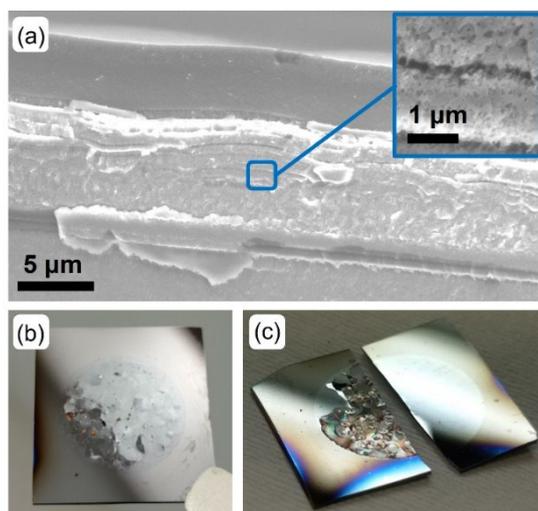

**Figure S2:** (a) SEM micrograph of a broken multilayer in which the LPL layer is etched using method A in non-ethanolic electrolyte ([HF]=50%), $j = 80$ mA/cm$^2$), (b) picture of the aforementioned sample and (c) picture of a detachable foil sample in which only the HPL is prepared using the same electrolyte and $j = 90$ mA/cm$^2$.

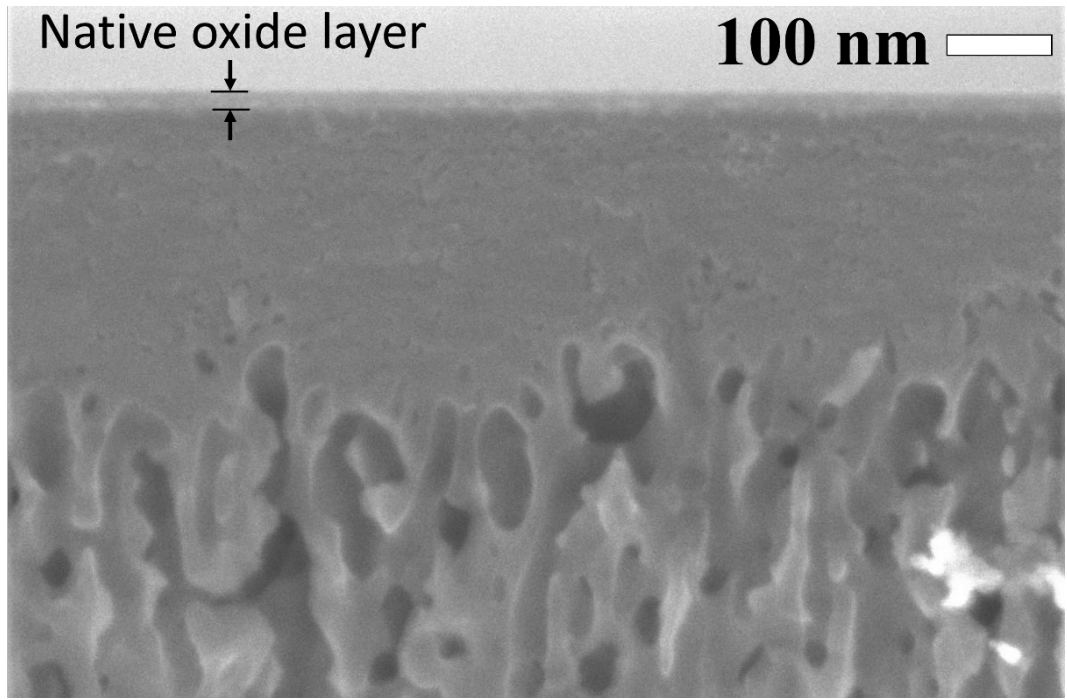

**Figure S3.** Illustrative cross-sectional SEM showing the top part and of etched plus sintered multilayer in which a 10-15 nm layer of native oxide is noticeable.

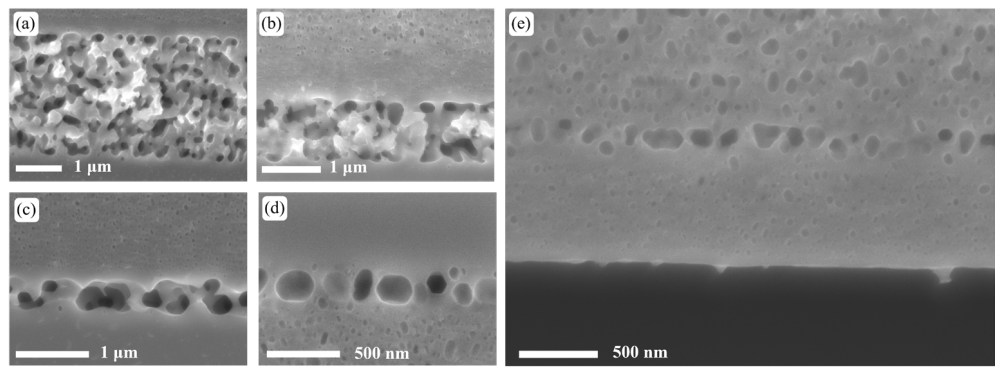

**Figure S4.** SEM micrographs of HPL pore structure evolution with varying HF concentration and etching time for (a) [HF]=25% and a given time $t$, (b) [HF]=17% and $t$, (c) [HF]=17% and $t/2$, (d) [HF]=17% and $t/3$ and (e) [HF]=12% and $t$.

**Table S1**. Average values and standard deviations of RMS measurements for foil surfaces and parent substrate surfaces of all samples.

| Sample ID | RMS values (nm) | |
|---|---|---|
| | Foil surface | Parent substrate surface |
| BL-AE | 0.33 (0.09) | - |
| BL-BE | 0.29 (0.02) | - |
| BL-AS | 0.78 (0.02) | - |
| BL-BS | 0.28 (0.01) | - |
| ML2-A | 2.12 (0.34) | 21.25 (1.42) |
| ML2-B | 1.11 (1.05) | 10.17 (1.65) |
| ML10-A | 1.85 (0.08) | 8.34 (1.60) |
| ML10-B | 0.89 (0.14) | 8.71 (2.37) |

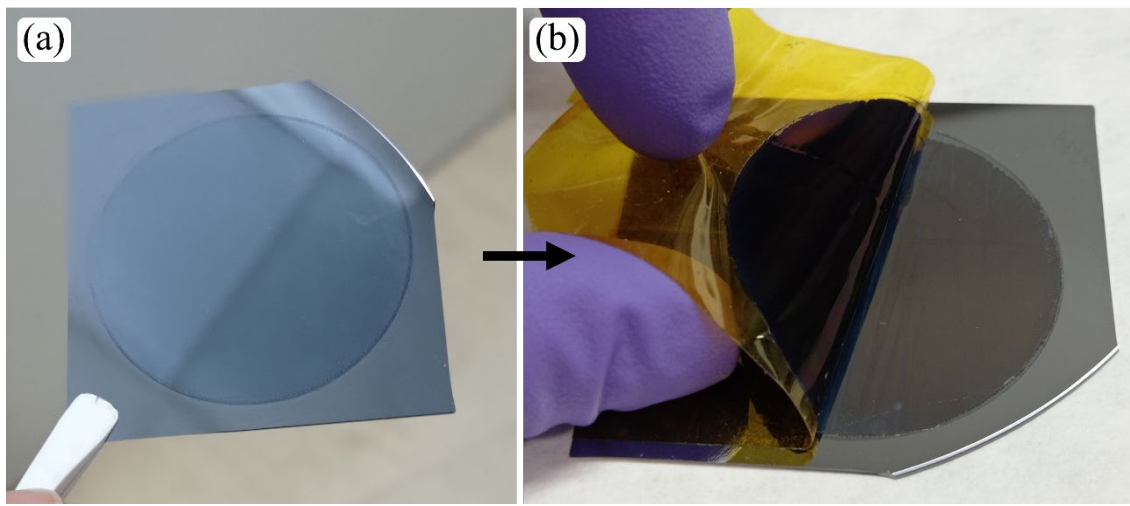

**Figure S5.** Pictures showing (a) an etched 10 μm multilayer over a 19.6 cm² area (circle diameter is 2") prepared *via* method B and (b) how it easily detaches from the parent substrate.